\documentstyle[aps,prb,epsfig]{revtex}
\begin{document}
\draft
\author{Mikio Eto$^{1}$ and Yuli V.\ Nazarov$^{2}$}
\address{$^1$Faculty of Science and Technology, Keio
University,
3-14-1 Hiyoshi, Kohoku-ku, Yokohama 223-8522, Japan \\
$^2$Department of Applied Physics/DIMES,
Delft University of Technology,
Lorentzweg 1, 2628 CJ Delft, The Netherlands}
\title{Multi-parameter scaling of the Kondo effect in quantum dots with
\\ an even number of electrons}
\date{March 7, 2002}
\maketitle
\begin{abstract}
We address a recent theoretical discrepancy 
concerning the Kondo effect in quantum dots with an
even number of electrons where spin-singlet and -triplet states are nearly
degenerate. We show that the discrepancy arises from the fact that
the Kondo scaling involves many parameters,
which makes the results depend on concrete microscopic models.
We illustrate this by the scaling calculations of the Kondo temperature,
$T_{\rm K}$, as a function of the energy difference between the singlet
and triplet states $\Delta$.
$T_{\rm K}(\Delta)$ decreases with increasing $\Delta$, showing a crossover
from a power law with a universal exponent to that with a nonuniversal exponent.
The crossover depends on the initial parameters of the model.
\end{abstract}

Recent observations of the Kondo effect in semiconductor quantum
dots have attracted much attention.\cite{kondo1} 
In quantum dots, the number of electrons $N$ is fixed by the Coulomb
blockade to integer values and can be tuned by gate voltages.
The usual Kondo effect takes place for an odd $N$. Here, the spin
$1/2$ is formed in the dot and it is coupled to the Fermi sea of
conduction electrons in external leads through tunneling barriers.

Strikingly the Kondo effect has also been observed for an even $N$
in some special cases.\cite{Sasaki,kondo4,nanotube}
In so-called ``vertical'' quantum dots, the ground state can be changed 
from a spin triplet to singlet by the magnetic field.\cite{Tarucha}
Sasaki {\it et al}.\ have found a pronounced Kondo effect near 
the degeneracy point between the spin states.\cite{Sasaki}
In this experiment the Zeeman splitting of the triplet state can
be neglected owing to a small $g$ factor in semiconductor
heterostructures.\cite{Zeeman}

In our previous paper, we have suggested an explanation of this phenomenon
within a model.\cite{me}
The Kondo temperature $T_{\rm K}$ has been evaluated as a function of
the energy difference between the singlet and triplet states,
$\Delta$, using the ``poor man's'' scaling method.
We have shown that (i) $T_{\rm K}(\Delta)$ is maximal around $\Delta=0$,
(ii) for positive $\Delta$, $T_{\rm K}(\Delta)$ decreases with
increasing $\Delta$ obeying a power law, $T_{\rm K}(\Delta)
\propto 1/\Delta^{\gamma}$, and (iii) for negative $\Delta$,
the Kondo effect is not relevant when $|\Delta| \gg T_{\rm K}(0)$.
The exponent $\gamma$ is not universal ($0<\gamma \le 1$) depending on a
ratio of the initial coupling constants. 
Our results have indicated an enhancement of the Kondo effect by
the competition between singlet and triplet states, qualitatively in
agreement with the experimental results by Sasaki {\it et al}.\cite{Sasaki}
After our work, Pustilnik and Glazman have considered a different
model for this ``triplet-singlet Kondo effect''.\cite{Pustilnik}
Their results are qualitatively the same as ours but quantitatively
different: They have derived a power law of $T_{\rm K}(\Delta)$ with
a universal exponent, $\gamma= 2+\sqrt{5}$.

The purpose of this brief report is to elucidate the discrepancy between
the results mentioned above. It should be noted that a microscopic model
sets the initial values of coupling constants, which are different between
in Refs.\ \onlinecite{me} and \onlinecite{Pustilnik}, whereas a
system of the scaling equations describes the evolution of the coupling
constants. If it were a one-parameter scaling, both models would result
in the identical results. However, this is not the present case of
{\it multi-parameter} scaling.
The scaling equations are
\begin{equation}
\left\{
\matrix{
d J^{(1)} / d \ln D & = & -2\nu \left[ J^{(1) 2} + (\tilde{J}_1^2 +
\tilde{J}_2^2 )/2 \right],
\cr
d J^{(2)} / d \ln D & = & -2\nu \left[ J^{(2) 2} + (\tilde{J}_1^2 +
\tilde{J}_2^2 )/2 \right],
\cr
d \tilde{J}_1/ d \ln D & = & -2\nu (J^{(1)}+J^{(2)})\tilde{J}_1 + \nu J'
\tilde{J}_1,
\cr
d \tilde{J}_2/ d \ln D & = & -2\nu (J^{(1)}+J^{(2)})\tilde{J}_2 - \nu J'
\tilde{J}_2,
\cr
d J'/ d \ln D & = & 8 \nu ( \tilde{J}_1^2 - \tilde{J}_2^2 ),
}
\right.
\label{eq:scalA}
\end{equation}
when the energy scale $D$ is much larger than $|\Delta|$.
This system of the equations encompasses five coupling constants
($J^{(1)},J^{(2)},\tilde{J}_1,\tilde{J}_2,J'$) and exhibits a complicated
non-linear behavior. Our previous results correspond to a
unstable fixed line of these equations, whereas the results of Ref.\
\onlinecite{Pustilnik} to a stable fixed line.
We show that in general the coupling constants develop along the
unstable fixed line at first and transit toward the stable fixed line
with decreasing $D$. The stable fixed line is much less relevant than
in the situation of one-parameter scaling.
Indeed our numerical calculations
indicate that it may take up to six decades to reach the
neighborhood of the stable fixed line (see Fig.\ 1(a)).
We find that $T_{\rm K}(\Delta)$ exhibits a crossover
between power laws with nonuniversal and universal exponents, which are
determined by the unstable and stable fixed lines, respectively.
Besides, we show that our previous calculations along the unstable
fixed line provide a lower limit of $T_{\rm K}(\Delta)$ in an
analytical form.

In a Coulomb blockade region with an even $N$, we consider
two electrons at the background of a singlet state of
the other $N-2$ electrons. We regard the background as the vacuum
$|0 \rangle$.
The two electrons occupy two levels of different orbital symmetry,
$\varepsilon_1, \varepsilon_2$, and make a spin-triplet state ($S=1$) or
-singlet state ($S=0$):
$\displaystyle
|S M \rangle = |1 1 \rangle = d_{1 \uparrow}^{\dagger} d_{2 \uparrow}^{\dagger}
|0 \rangle$,
$\displaystyle
|1 0 \rangle = (1/\sqrt{2})
(d_{1 \uparrow}^{\dagger} d_{2 \downarrow}^{\dagger}
+d_{1 \downarrow}^{\dagger} d_{2 \uparrow}^{\dagger}) |0 \rangle$,
$\displaystyle
|1 -1 \rangle = d_{1 \downarrow}^{\dagger} d_{2 \downarrow}^{\dagger}
|0 \rangle$,
$\displaystyle
|0 0 \rangle = (1/\sqrt{2})
(C_1 d_{1 \uparrow}^{\dagger} d_{1 \downarrow}^{\dagger}
 -C_2 d_{2 \uparrow}^{\dagger} d_{2 \downarrow}^{\dagger}) |0 \rangle$,
where $d_{i \sigma}^{\dagger}$ creates an electron with spin $\sigma$
in level $i$. The energy difference between the spin states is denoted by
$\Delta=E_{S=0}-E_{S=1}$.
The coefficients in the singlet state, $C_1$ and $C_2$ ($|C_1|^2+|C_2|^2=2$),
should be determined by the electron-electron interaction
and one-electron level spacing $\delta=\varepsilon_2-\varepsilon_1$.
We have set $C_1=C_2$ in our previous work,\cite{me} whereas $C_2=0$ in Ref.\
\onlinecite{Pustilnik}. Now we investigate general cases with respect to
$C_1$ and $C_2$. The other singlet states with higher energies are disregarded.

The level $i$ ($=1,2$) in the dot is connected to two external leads, $L$, $R$,
by the tunnel coupling, $V_{L,i}$, $V_{R,i}$. We assume two
channels in the leads; the orbital symmetry is conserved in the tunneling
processes.\cite{com1} We perform the unitary transformation for the conduction
electrons in the two leads,\cite{Glazman}
$c_{k \sigma}^{(i)}=(V_{L,i}^*c_{L,k \sigma}^{(i)}+
V_{R,i}^*c_{R,k \sigma}^{(i)})/V_i$,
$\bar{c}_{k \sigma}^{(i)}=(-V_{R,i}c_{L,k \sigma}^{(i)}+
V_{L,i}c_{R,k \sigma}^{(i)})/V_i$ with
$V_i=\sqrt{|V_{L,i}|^2+|V_{R,i}|^2}$, where $c_{\alpha,k\sigma}^{(i)}$
is the annihilation operator of an electron in lead $\alpha$,
with momentum $k$, spin $\sigma$, and orbital symmetry $i$ $(=1,2)$.
The mode $c_{k \sigma}^{(i)}$ is coupled to the dot with $V_i$. The mode
$\bar{c}_{k \sigma}^{(i)}$ is decoupled from the lead, which shall be disregarded.
The Hamiltonian of the leads and that of tunneling processes are written as
\begin{eqnarray}
H_{\rm{leads}} & = & \sum_{k \sigma i}
\varepsilon_{k}^{(i)} c_{k\sigma}^{(i) \dagger}
c_{k\sigma}^{(i)}, \\
H_{\rm T} & = & \sum_{k \sigma i} V_{i}
(c_{k\sigma}^{(i) \dagger} d_{i \sigma} + \rm{H.c.}),
\end{eqnarray}
respectively.
The density of states $\nu$ in the leads remains constant in the
energy band of $[-D, D]$.

In the Coulomb blockade region,
the addition and extraction energies,
$E^{\pm} \equiv E(N \pm 1) -E(N) \mp \mu$, are much larger than
the level broadening $\Gamma_i=\pi\nu V_i^2$ ($i=1,2$) and temperature $T$,
where $E(N)$ is the energy with $N$ electrons in the dot and
$\mu$ is the Fermi energy in the leads.
We also assume that $E^{\pm} \gg |\Delta|, \delta$.
In the similar way to Ref.\ \onlinecite{me}, we integrate out the states
with $N \pm 1$ electrons and obtain the effective low-energy
Hamiltonian,
\begin{equation}
H_{\rm{eff}}=H_{\rm{leads}}+H_{\rm{dot}}+
H^{S=1}+H^{S=1 \leftrightarrow 0}+H_{\rm{eff}}^{\prime}.
\end{equation}
The Hamiltonian of the dot, $H_{\rm{dot}}$, reads
\begin{equation}
H_{\rm{dot}}=\sum_{S,M} E_{S} f_{SM}^{\dagger}f_{SM},
\end{equation}
using pseudofermion operators $f_{SM}^{\dagger}$ ($f_{SM}$) which
create (annihilate) the state $|SM \rangle$. It is required that
$\displaystyle \sum_{SM} f_{SM}^{\dagger} f_{SM} =1$.
$H^{S=1}$ describes the spin-flip processes among three components of
the triplet state,
\begin{eqnarray}
H^{S=1} =
\sum_{k k'} \sum_{i=1,2} J^{(i)}
\Bigl[ \sqrt{2}(f_{11}^{\dagger}f_{10}+f_{10}^{\dagger}f_{1 -1})
                  c_{k' \downarrow}^{(i) \dagger} c_{k \uparrow}^{(i)}
+\sqrt{2}(f_{10}^{\dagger}f_{11}+f_{1 -1}^{\dagger}f_{10})
                  c_{k' \uparrow}^{(i) \dagger} c_{k \downarrow}^{(i)}
            \nonumber \\
 +(f_{11}^{\dagger}f_{11}-f_{1 -1}^{\dagger}f_{1 -1})
                  (c_{k' \uparrow}^{(i) \dagger} c_{k \uparrow}^{(i)}
               -c_{k' \downarrow}^{(i) \dagger} c_{k \downarrow}^{(i)})
\Bigr].
\label{eq:H1}
\end{eqnarray}
The exchange couplings are
$J^{(i)} = V_{i}^2/(2E_{\rm c})$ ($i=1,2$), 
with $1/E_{\rm c}=1/E^+ +1/E^-$, which are accompanied by the scattering
of a conduction electron of channel $i$.
$H^{S=1 \leftrightarrow 0}$ represents the conversion between the triplet
and singlet states, 
with the interchannel scattering of a conduction electron,
\begin{eqnarray}
H^{S=1 \leftrightarrow 0} =
\sum_{k k'}
\Bigl[\sqrt{2}(\tilde{J}_1 f_{11}^{\dagger}f_{00}
              -\tilde{J}_2 f_{00}^{\dagger}f_{1 -1})
                 c_{k' \downarrow}^{(1) \dagger} c_{k \uparrow}^{(2)}
+\sqrt{2}(\tilde{J}_2 f_{00}^{\dagger}f_{11}
         -\tilde{J}_1 f_{1 -1}^{\dagger}f_{00})
                 c_{k' \uparrow}^{(1) \dagger} c_{k \downarrow}^{(2)}
\nonumber \\
-(\tilde{J}_1 f_{10}^{\dagger}f_{00}+\tilde{J}_2 f_{00}^{\dagger}f_{10})
                 (c_{k' \uparrow}^{(1) \dagger} c_{k \uparrow}^{(2)}
                 -c_{k' \downarrow}^{(1) \dagger} c_{k \downarrow}^{(2)})
+ (1 \leftrightarrow 2)  \Bigr],
\end{eqnarray}
where $\tilde{J}_i = C_i V_{1} V_{2}/(2E_{\rm c})$.
The last term of $H_{\rm{eff}}$ represents the scattering processes without
spin-flip in the dot,
\begin{equation}
H_{\rm{eff}}'=
\sum_{k k' \sigma}\sum_{i=1,2}
  \left[ J^{\prime (i)} c_{k' \sigma}^{(i) \dagger} c_{k \sigma}^{(i)}
            \sum_M f_{1 M}^{\dagger}f_{1 M}
              + J^{\prime\prime (i)} c_{k' \sigma}^{(i) \dagger}
               c_{k \sigma}^{(i)} f_{00}^{\dagger}f_{00}
        \right].
\end{equation}

We calculate $T_{\rm K}$ using the poor man's scaling
method.\cite{Anderson}
With changing the energy scale (bandwidth of the conduction electrons)
from $D$ to $D-|d D|$, we renormalize
the exchange couplings not to change the low-energy physics,
within the second-order perturbation with respect
to $H^{S=1}+H^{S=1 \leftrightarrow 0}+H_{\rm{eff}}^{\prime}$.
This procedure yields the scaling equations in two limits.
For $D \gg |\Delta|$, $H_{\rm{dot}}$ can be safely disregarded
in $H_{\rm{eff}}$. The scaling equations are given by Eqs.\
(\ref{eq:scalA}) with
$J'=J^{\prime (1)}-J^{\prime (2)}-J^{\prime\prime (1)}+J^{\prime\prime
(2)}$.
For $D \ll \Delta$, the ground state of the dot is a spin triplet
and the singlet state can be disregarded. Then $J^{(1)}$ and $J^{(2)}$
evolve independently,
\begin{equation}
\frac{d}{d\ln D} J^{(i)} = -2\nu J^{(i) 2},
\label{eq:scalB}
\end{equation}
whereas the other coupling constants do not change.
The Kondo temperature is determined as the energy scale at which
the coupling constants become so large that the perturbation
breaks down.

In our previous model,\cite{me} $\tilde{J}_1=\tilde{J}_2$ and the scaling
equations (\ref{eq:scalA}) are closed without $J'$.
The ratio of $J^{(2)}/J^{(1)}$ is also fixed in the scaling by
Eqs.\ (\ref{eq:scalA}). We obtain the Kondo temperature as a function of
$\Delta$, as follows.
(i) When $|\Delta| \ll T_{\rm K}(0)$, the scaling equations
(\ref{eq:scalA}) remain valid till the scaling ends. This yields
\begin{equation}
T_{\rm K}(0) = D_0 \exp [-1/2\nu (J^{(1)}+J^{(2)})],
\label{eq:TK0a}
\end{equation}
where $D_0$ is the initial bandwidth, which is given by
$\sqrt{E^+ E^-}$.\cite{Haldane}
(ii) When $\Delta > D_0$, the scaling equations (\ref{eq:scalB}) work
in the whole scaling region. $T_{\rm K}$ is
identical to that of a localized spin with $S=1$,\cite{Okada}
\begin{equation}
T_{\rm K}(\infty) = D_0 \exp [-1/2\nu J^{(1)}],
\label{eq:TK0b}
\end{equation}
when $J^{(1)} \ge J^{(2)}$. (iii) 
In the intermediate region of $T_{\rm K}(0) \ll \Delta \ll D_0$, we
match the solutions of Eqs.\ (\ref{eq:scalA}) and (\ref{eq:scalB}) at
$D \simeq \Delta$ and obtain a power law
\begin{equation}
T_{\rm K}(\Delta) = T_{\rm K}(0)\cdot 
\left( T_{\rm K}(0)/\Delta \right)^{\gamma}.
\label{eq:TK0c}
\end{equation}
The exponent is given by $\gamma=J^{(2)}/J^{(1)}$.
(iv) For $\Delta <0$, $T_{\rm K}$ drops to zero suddenly at
$\Delta \sim -T_{\rm K}(0)$.

In general situations of $C_1 \ne C_2$, we find that the line of
$J^{(2)}/J^{(1)}=$ const.\ and $\tilde{J}_1=\tilde{J}_2$ is not stable
in the scaling by Eqs.\ (\ref{eq:scalA}). The renormalization flow follows
this line at the beginning, but finally goes to a fixed point of
$J^{(1)}=J^{(2)}=\infty$, $J^{(2)}/J^{(1)}=1$, 
$\tilde{J}_1/J^{(1)}=\sqrt{2(2+\sqrt{5})}$, $\tilde{J}_2/J^{(1)}=0$,
and $J'/J^{(1)}=-2(1+\sqrt{5})$, as the energy scale $D$
decreases to $T_{\rm K}$. Around this fixed point,
the stable fixed line discussed in Ref.\ \onlinecite{Pustilnik} is
relevant. By expanding the coupling constants around
the fixed point to the first order of $1/\ln (D/T_{\rm K})$,
Pustilnik and Glazman have obtained a power law of
$T_{\rm K}(\Delta)$ with a universal exponent,
$\gamma= 2+\sqrt{5}$.\cite{Pustilnik} However, the range of energy scale
is very limited where the physical properties are determined by the
stable fixed line, due to the complexity of the multi-parameter scaling.

To elucidate this multi-parameter scaling,
we solve the scaling equations numerically when $\Delta \ge 0$;
Eqs.\ (\ref{eq:scalA}) for $D>\Delta$ and Eqs.\ (\ref{eq:scalB})
for $D<\Delta$.
The Kondo temperature $T_{\rm K}$ is determined as the energy
scale $D$ at which $\nu (J^{(1)}+J^{(2)}) = 1$.
First we examine the scaling by Eqs.\ (\ref{eq:scalA}),
assuming $\Delta=0$. In Fig.\ 1(a),
we show the ratio of $(J^{(1)}-J^{(2)})/(J^{(1)}+J^{(2)})$
as a function of $\log D/T_{\rm K}(0)$, where $T_{\rm K}(0)$ is the
Kondo temperature at $\Delta=0$. We set $J^{(2)}/J^{(1)}$
to be {\it a}, $1$; {\it b}, $0.5$; and {\it c}, $0.3$ in the initial
condition.\cite{com2}
In the case of $C_1=C_2$ (solid lines), the ratio is fixed at a
nonuniversal value determined by the given condition.
In the cases of $C_2=0$ (broken lines) and $C_2/C_1=0.6$ (dotted lines),
the ratio gradually deviates from the initial value with decreasing $D$,
and finally goes to zero.
This limit corresponds to the fixed point of the renormalization,
$J^{(2)}/J^{(1)}=1$. It may need to reduce the energy scale by six decades
to reach its vicinity.
(The case {\it a} with $J^{(1)}=J^{(2)}$ is an exception,
where the fixed point is on the line of $J^{(2)}/J^{(1)}=$const.)
Similarly Fig.\ 1(b) indicates $(\tilde{J}_1-\tilde{J}_2)/(J^{(1)}+J^{(2)})$.
In the case of $C_1=C_2$, the ratio is fixed at zero
($\tilde{J}_1=\tilde{J}_2$). Otherwise, the ratio goes to a value
of the fixed point, $\sqrt{2+\sqrt{5}}/\sqrt{2}$,
as $D$ decreases.

Figure 2 presents the Kondo temperature as a function
of $\Delta$ on a log-log scale;
{\it a}, $J^{(2)}/J^{(1)}=1$ and {\it b}, $0.3$.
Both $T_{\rm K}$ and $\Delta$ are normalized by $T_{\rm K}(0)$.
In all the cases, the Kondo temperature increases with decreasing
$\Delta$.
In the case of $C_1=C_2$ (solid lines), $T_{\rm K}(\Delta)/T_{\rm K}(0)$
obeys a power law, Eq.\ (\ref{eq:TK0c}), with
a nonuniversal exponent, $\gamma=J^{(2)}/J^{(1)}$.
In the cases of $C_2=0$ (broken lines) and $C_2/C_1=0.6$ (dotted lines),
$T_{\rm K}(\Delta)$ shows a crossover from the power law with
$\gamma=J^{(2)}/J^{(1)}$ at large $\Delta/T_{\rm K}(0)$ to that with
$\gamma=2+\sqrt{5} \approx 4.2$ at small $\Delta/T_{\rm K}(0)$.
The crossover depends on the initial values of $C_2/C_1$ and
$J^{(2)}/J^{(1)}$.
The universal exponent is seen in quite limited situations.

Finally we compare the Kondo temperature with various $C_2/C_1$ and
$J^{(2)}/J^{(1)}$ in the initial condition. We choose
$\nu (J^{(1)}+J^{(2)}) = 0.1$ which corresponds to the experimental
situation with $E^{\pm} \approx$ 5K and level broadening $\Gamma \approx$
1.5K.\cite{Sasaki} In Fig.\ 3, we show $T_{\rm K}$ in
units of $D_0 \exp [-1/2\nu (J^{(1)}+J^{(2)})]$ (which is identical to
$T_{\rm K}(0)$ with $C_1=C_2$, Eq.\ (\ref{eq:TK0a})), as a function of
$\Delta$.
We find that the case with $C_1=C_2$ (solid lines) provides a lower
limit of $T_{\rm K}$ for each value of $J^{(2)}/J^{(1)}$
({\it a}, $1$; {\it b}, $0.5$; {\it c}, $0.3$),
whereas the case with $C_2=0$ (broken lines) gives a upper limit.
The inset to Fig.\ 3 shows $T_{\rm K}(0)$ as a function of
$J^{(2)}/J^{(1)}$ ($\equiv \tan^2\theta$).
For fixed $C_2/C_1$, $T_{\rm K}(0)$ is maximal at $J^{(2)}/J^{(1)}=1$
($\theta/\pi=0.25$) and decreases with decreasing $J^{(2)}/J^{(1)}$.
In a case of $C_2=0$ and $J^{(2)}/J^{(1)}=1$, the development of the
coupling constants by Eqs.\ (\ref{eq:scalA})
is restricted in a subspace with $\tilde{J}_2(D)=0$
and $J^{(1)}(D)=J^{(2)}(D)$, and in consequence they reach the fixed
point fastest. This results in the highest $T_{\rm K}$ for a given
value of $J^{(1)}+J^{(2)}$.
As the initial values are deviated from the condition,
the renormalization flow to the fixed point needs more ``time'' in a
larger space. (If $D$ reaches $\Delta$ before $T_{\rm K}$, the coupling
constants follow Eqs.\ (\ref{eq:scalB}) at $T_{\rm K}<D<\Delta$,
which is common to all the cases.)
Since our previous model with $C_1=C_2$ is the farthest
from the condition of $C_2=0$, it yields a lower limit of the Kondo
temperature.

In conclusion,
we have examined a generalized model for the Kondo effect
which involves spin-singlet and -triplet states in quantum dots,
to elucidate the discrepancy between the previous studies.\cite{me,Pustilnik}
The Kondo temperature $T_{\rm K}$ is calculated as a function of
the energy difference between the states, $\Delta$, using the
scaling method.
The function of $T_{\rm K} (\Delta)$
shows a crossover between power laws with a nonuniversal exponent
($\gamma=J^{(2)}/J^{(1)}$) and with a universal exponent
($\gamma=2+\sqrt{5}$).
Although our previous model with $\tilde{J}_1=\tilde{J}_2$
($C_1=C_2$ in the singlet state)\cite{me} is not complete
to discuss the triplet-singlet Kondo effect, it yields a lower limit
of $T_{\rm K} (\Delta)$ in an analytical form,
Eqs.\ (\ref{eq:TK0a}), (\ref{eq:TK0b}), and (\ref{eq:TK0c}).

The authors gratefully acknowledge discussions with
L.\ P.\ Kouwenhoven, S.\ De Franceschi,
J.\ M.\ Elzerman, K.\ Maijala, S.\ Sasaki,
W.\ G.\ van der Wiel, Y.\ Tokura, L.\ I.\ Glazman, M.\ Pustilnik,
T.\ Saso, and G.\ E.\ W.\ Bauer.

\pagebreak

\noindent
{\large Figure captions}
\vspace{.5cm}

Fig.\ 1:
The scaling of the coupling constants by
Eqs.\ (\ref{eq:scalA}), assuming $\Delta=0$.
(a) $(J^{(1)}-J^{(2)})/(J^{(1)}+J^{(2)})$ and (b)
$(\tilde{J}_1-\tilde{J}_2)/(J^{(1)}+J^{(2)})$
as functions of $\log D/T_{\rm K}(0)$, where $T_{\rm K}(0)$
is the Kondo temperature at $\Delta=0$.
$J^{(2)}/J^{(1)}$ is {\it a}, $1$; {\it b}, $0.5$;
and {\it c}, $0.3$. 
The cases with $C_1=C_2$ are drawn by solid lines,
$C_2=0$ by broken lines, and $C_2/C_1=0.6$ by dotted lines.
Note that three lines are overlapped for
case {\it a} in (a).
Initially, we choose $\nu (J^{(1)}+J^{(2)}) = 0.01$.

\vspace{0.5cm}

Fig.\ 2:
The Kondo temperature, $T_{\rm K}/T_{\rm K}(0)$, as a function
of $\Delta/T_{\rm K}(0)$, on a log-log scale ($\Delta>0$).
$T_{\rm K}(0)$ is the Kondo temperature at $\Delta=0$.
$J^{(2)}/J^{(1)}$ is {\it a}, $1$ and {\it b}, $0.3$. 
The cases with $C_1=C_2$ are drawn by solid lines,
$C_2=0$ by broken lines, and $C_2/C_1=0.6$ by dotted lines.
Initially, we choose $\nu (J^{(1)}+J^{(2)}) = 0.01$.

\vspace{0.5cm}

Fig.\ 3:
The Kondo temperature $T_{\rm K}$ as a function of $\Delta$ ($\Delta>0$).
The units of $T_{\rm K}$ is $D_0 \exp [-1/2\nu (J^{(1)}+J^{(2)})]$
where $D_0$ is the initial bandwidth ($D_0=\sqrt{E^+ E^-} \approx 5$K
in the experiment of Ref.\ \onlinecite{Sasaki}).
$J^{(2)}/J^{(1)}$ is {\it a}, $1$; {\it b}, $0.5$;
and {\it c}, $0.3$. 
The cases with $C_1=C_2$ are drawn by solid lines,
$C_2=0$ by broken lines, and $C_2/C_1=0.6$ by dotted lines.
As the initial condition, we choose $\nu (J^{(1)}+J^{(2)}) = 0.1$.
Inset: The Kondo temperature at $\Delta=0$, $T_{\rm K}(0)$, 
as a function of $\theta$, where $\tan^2\theta=J^{(2)}/J^{(1)}$.
$\theta/\pi=0.25$ ($0.20$, $0.15$) for $J^{(2)}/J^{(1)}=1$
($0.5$, $0.3$).

\pagebreak

\vspace*{2cm}
\epsfig{file=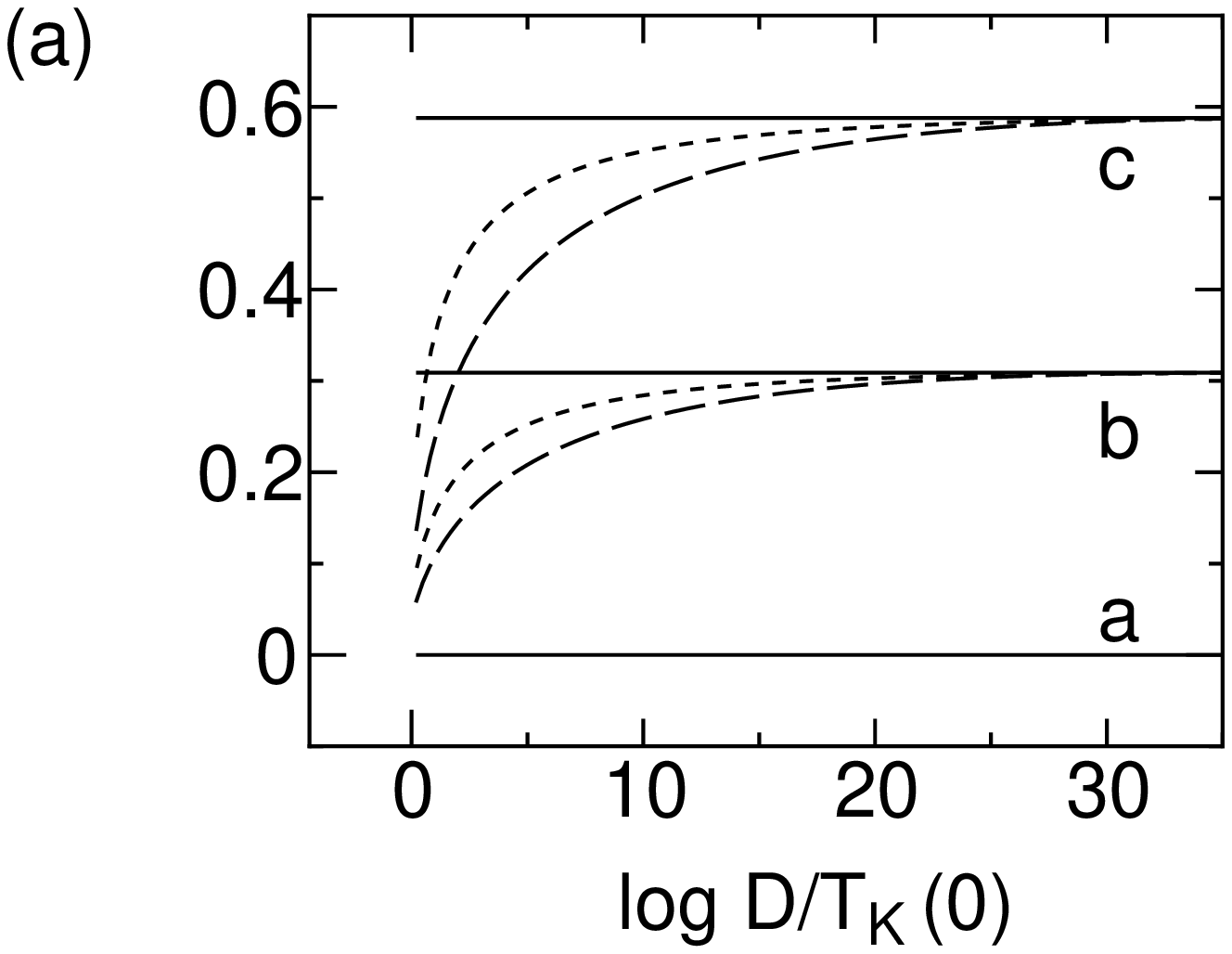,width=8cm}

\epsfig{file=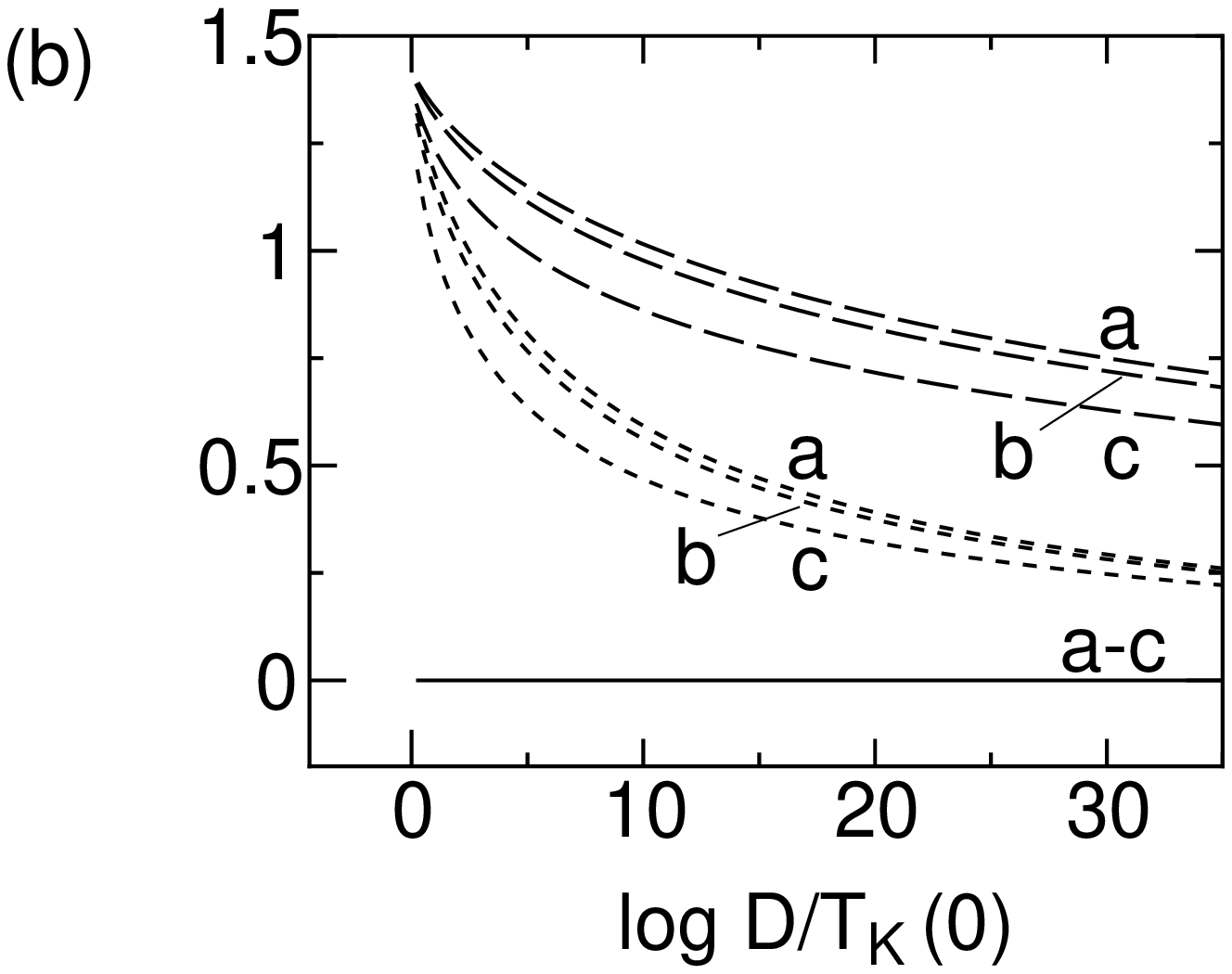,width=8cm}

  \vspace*{1cm}

\large{Figure 1 (M.\ Eto and Yu.\ V.\ Nazarov)}

\pagebreak

\epsfig{file=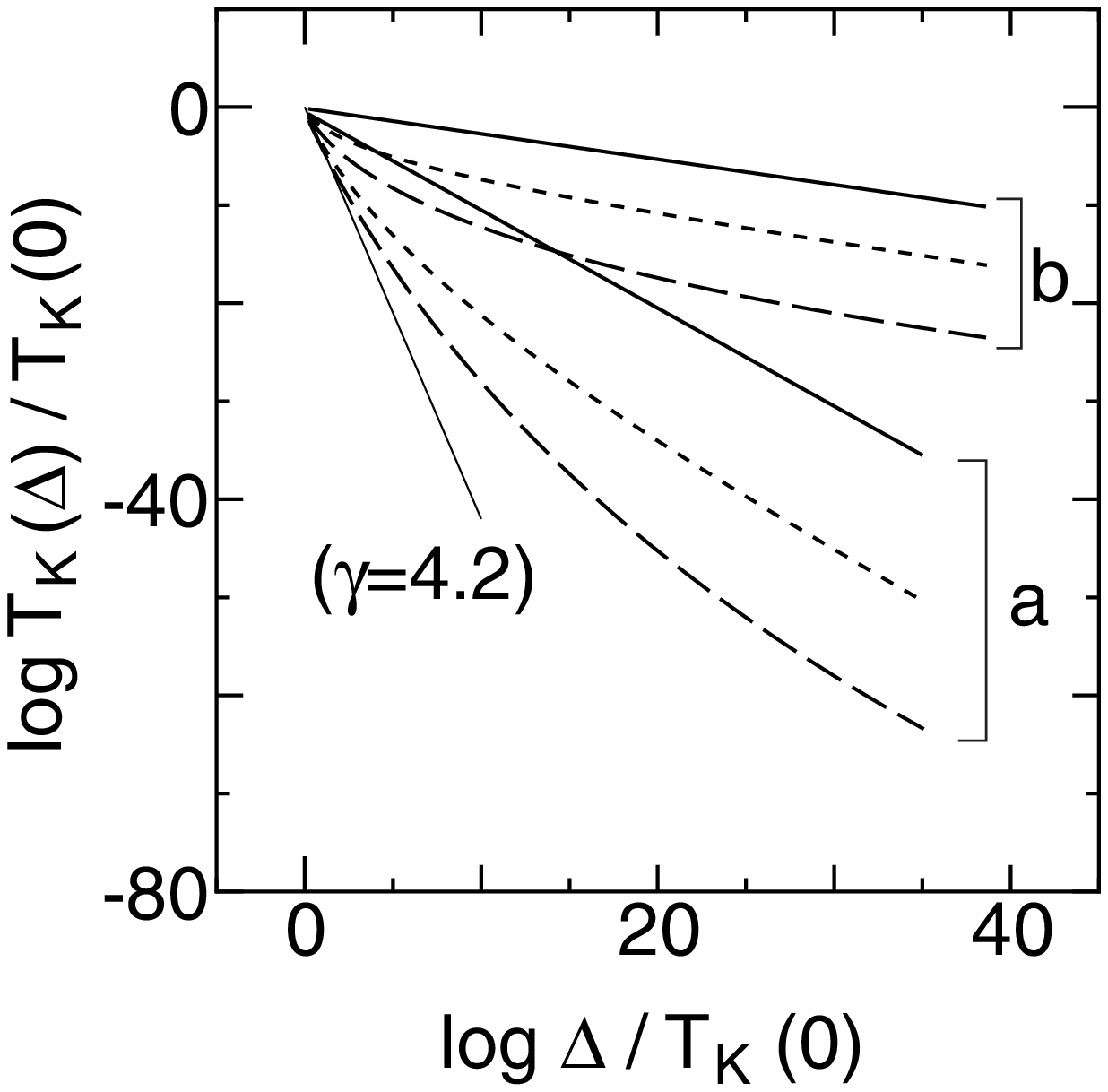,width=10cm}

  \vspace*{1cm}

\large{Figure 2 (M.\ Eto and Yu.\ V.\ Nazarov)}

\pagebreak

\epsfig{file=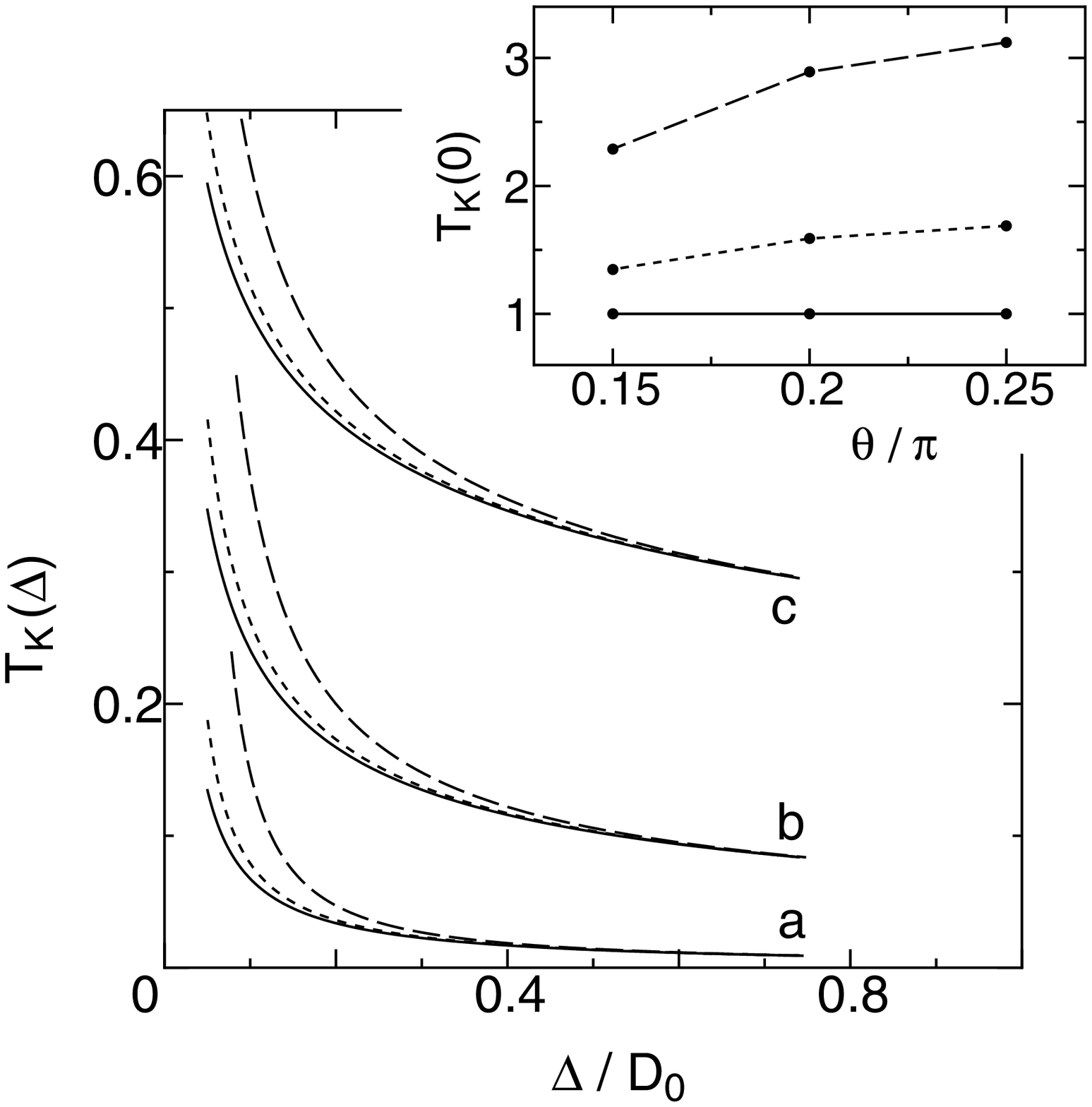,width=10cm}

  \vspace*{1cm}

\large{Figure 3 (M.\ Eto and Yu.\ V.\ Nazarov)}


\begin{references}
\bibitem{com1}
This is the case of vertical quantum dots in which 
the tunneling processes conserve the symmetry for the motion of electrons
in the transverse direction.
\end{references}
\end{document}